\documentstyle[12pt,aasms4,flushrt]{article}

\def\gsim{\;\rlap{\lower 2.5pt
\hbox{$\sim$}}\raise 1.5pt\hbox{$>$}\;}
\def\lsim{\;\rlap{\lower 2.5pt
\hbox{$\sim$}}\raise 1.5pt\hbox{$<$}\;}
\newcommand\beq{\begin{equation}}
\newcommand\eeq{\end{equation}}

\def\v{\vspace{-0.1in}}

\begin{document}
\Large 
\centerline{\bf Formation of the First Stars and Galaxies}
\normalsize 
\author{\bf Volker Bromm, Naoki Yoshida, Lars Hernquist \& Christopher F. McKee}
\noindent
\normalsize{$^{1}$Astronomy Department, University of Texas, 2511 Speedway, Austin, TX}\\ 
\normalsize{$^{2}$Institute for the Physics and Mathematics of the Universe, University of Tokyo, Kashiwa, Chiba 277-8568, Japan}\\
\normalsize{$^{3}$Harvard-Smithsonian Center for Astrophysics, 60 Garden Street, Cambridge, MA}\\
\normalsize{$^{4}$Departments of Physics and Astronomy, University of California, Berkeley, CA}\\
\vskip 0.2in 
\hrule 
\vskip 0.2in 
\noindent

{\bf Observations made using large ground-based and space-borne
telescopes have probed cosmic history all the way from the present-day
to a time when the Universe was less than a tenth of its present age.
Earlier on lies the remaining frontier, where the first stars, galaxies,
and massive black holes formed. They fundamentally transformed the early
Universe by endowing it with the first sources of light and chemical
elements beyond the primordial hydrogen and helium produced in the Big Bang.
The interplay of theory and upcoming observations promises to answer the
key open questions in this emerging field.}
\newline

The formation of the first stars and galaxies at the end
of the cosmic dark ages is one of the central problems in modern 
cosmology\cite{BarL01}$^-$\cite{SaasFe08}. It is thought that during
this epoch the Universe was transformed from its simple initial state
into a complex, hierarchical system, through the growth of structure
in the dark matter, by the input of heavy elements from the first stars,
and by energy injection from these stars and from the first black
holes\cite{BrommL04a}$^,$\cite{CiardiF05}. An important milestone
in our understanding was
reached after the introduction of the now standard cold dark matter (CDM)
model of cosmic evolution which posits that structure grew hierarchically,
such that small objects formed first and then merged to form increasingly larger
systems\cite{BFPR85}. Within this model, dark matter `minihalos' (see below),
forming a few hundred million years after the Big Bang,
were identified as the sites where the first stars formed\cite{CR86}.
Building on this general framework, and relying on the development
of efficient new computational tools, the fragmentation properties of
primordial gas inside such minihalos were investigated with numerical
simulations, leading to the result that the first stars, so-called
Population~III (Pop~III), were predominantly very
massive\cite{BrommL04a}$^,$\cite{Glover05} (see the Box for the 
terminology used in this review). Recently, the frontier has progressed to
the next step in the hierarchical build-up of structure, to the emergence
of the first galaxies whose formation took place after the first stars
had formed and affected their common environment. It is very timely to
review our current understanding and remaining challenges, since we are
just entering an exciting period of discovery, where new observational probes
are becoming available, and where advances in supercomputer technology
enable ever more realistic theoretical predictions.

We begin with the formation of the first stars, discussing the
physics underlying the prediction that they were very massive, and 
how this picture would be modified if the dark matter exhibited non-standard
properties on small scales. We next address the feedback effects from the
first stars, with one main result being that such feedback might 
delay 
subsequent star formation by up to $\sim 10^8$~years. Proceeding to the
assembly of the first galaxies, we discuss the important role of turbulence
and supernova (SN) feedback
during their formation. Intriguingly, the cold accretion streams that 
feed the turbulence in the centers 
of the primordial galaxies are
reminiscent of the recently proposed new paradigm for galaxy formation,
in which
such cold streams are invoked to explain the build-up of massive
galaxies at more recent cosmic times in a smooth, rather than merger-driven,
fashion\cite{Dekel09}. We conclude with
an outlook into the likely key developments over the next decade.\\

\noindent{\Large \bf Formation of the first stars}

While dark-matter halos can originate through the action of gravity
alone, the formation of luminous objects, such as stars and galaxies,
is a much more complicated process. For star formation to begin, a
sufficient amount of cold dense gas must accumulate in a dark halo.
In the early Universe, the primordial gas cannot efficiently cool
radiatively because atoms have excitation energies that are too high,
and molecules, which have accessible rotational energies, are very
rare. Trace amounts of molecular hydrogen (${\rm H}_2$) can be
produced via a sequence of reactions, ${\rm H} + e^- \rightarrow {\rm
H}^- + \gamma$, followed by ${\rm H}^- + {\rm H} \rightarrow {\rm H}_2
+ e^- $, and under the proper conditions this allows the gas to cool
and eventually condense to form stars\cite{GP98}.

Numerical simulations\cite{Abel02}$^-$\cite{Yoshida06} starting from
cosmological initial conditions show that primordial gas clouds 
formed in dark matter halos with virial temperature $T_{\rm vir} \sim
1,000$~K and mass $\sim 10^6 M_{\odot}$ (so-called `minihalos').
In the standard CDM model, the minihalos that 
were
the first sites
for star formation are expected to be in place at $z\sim 20-30$, when
the age of the Universe was just a few hundred million
years\cite{Teg97}.
These systems correspond to $3-4\,\sigma$ peaks in the cosmic density
field, which is statistically described as a Gaussian random field.
Such high-density peaks are expected to be strongly
clustered\cite{Gao07}, and thus feedback effects from the first stars
are important in determining the fate of the surrounding primordial
gas clouds.  It is very likely that only one star can be formed within
a gas cloud, because the far UV radiation from a single massive star
is sufficient to destroy all the H$_2$ in the parent gas
cloud\cite{Omukai99}$^,$\cite{Kitayama04}. In principle, a cloud that
forms one of the first stars could fragment into a binary or multiple
star system\cite{Machida07}$^,$\cite{Clark08}, but simulations based
on self-consistent cosmological initial conditions do not show
this\cite{YOH08}.  Although the exact number of stars per cloud cannot
be easily determined, the number is expected to be small, so that
minihalos will not be galaxies (see Box).

Primordial gas clouds undergo runaway collapse when sufficient mass is
accumulated at the center of a minihalo. The minimum mass at the
onset of collapse is determined by the Jeans mass (more precisely, the
Bonnor-Ebert mass), which can be written as
\begin{equation}
M_{\rm J} \simeq 500 M_{\odot} \left(\frac{T}{200 {\rm \,K}}\right)^{3/2} 
\left(\frac{n}{10^4 {\rm \,cm}^{-3}}\right)^{-1/2} 
\end{equation}
for an atomic gas with temperature $T$ and particle number density
$n$.  The characteristic temperature is set by the energy separation
of the lowest-lying rotational levels of the trace amounts of H$_{2}$,
and the characteristic density corresponds to the thermalisation of
these levels, above which cooling becomes less
efficient\cite{Bromm02}.
A number of atomic and molecular processes are involved in the
subsequent evolution of a gravitationally collapsing gas.  It has been
suggested that a complex interplay between chemistry, radiative
cooling and hydrodynamics leads to fragmentation of the
cloud\cite{Silk83}, but vigorous fragmentation is
not observed even in extremely high resolution cosmological
simulations\cite{Abel02}$^-$\cite{Yoshida06}$^,$\cite{YOH08}$^,$\cite{BLoeb04}.
Interestingly, however, simulations starting from non-cosmological
initial conditions have yielded multiple cloud
cores\cite{Clark08}$^,$\cite{Bromm99}. It appears that a high initial
degree of spin in the gas eventually leads to the formation of a disk
and its subsequent break-up.  It remains to be seen whether such
conditions occur from realistic cosmological initial conditions.

\begin{figure*}[hptb]
\plotone{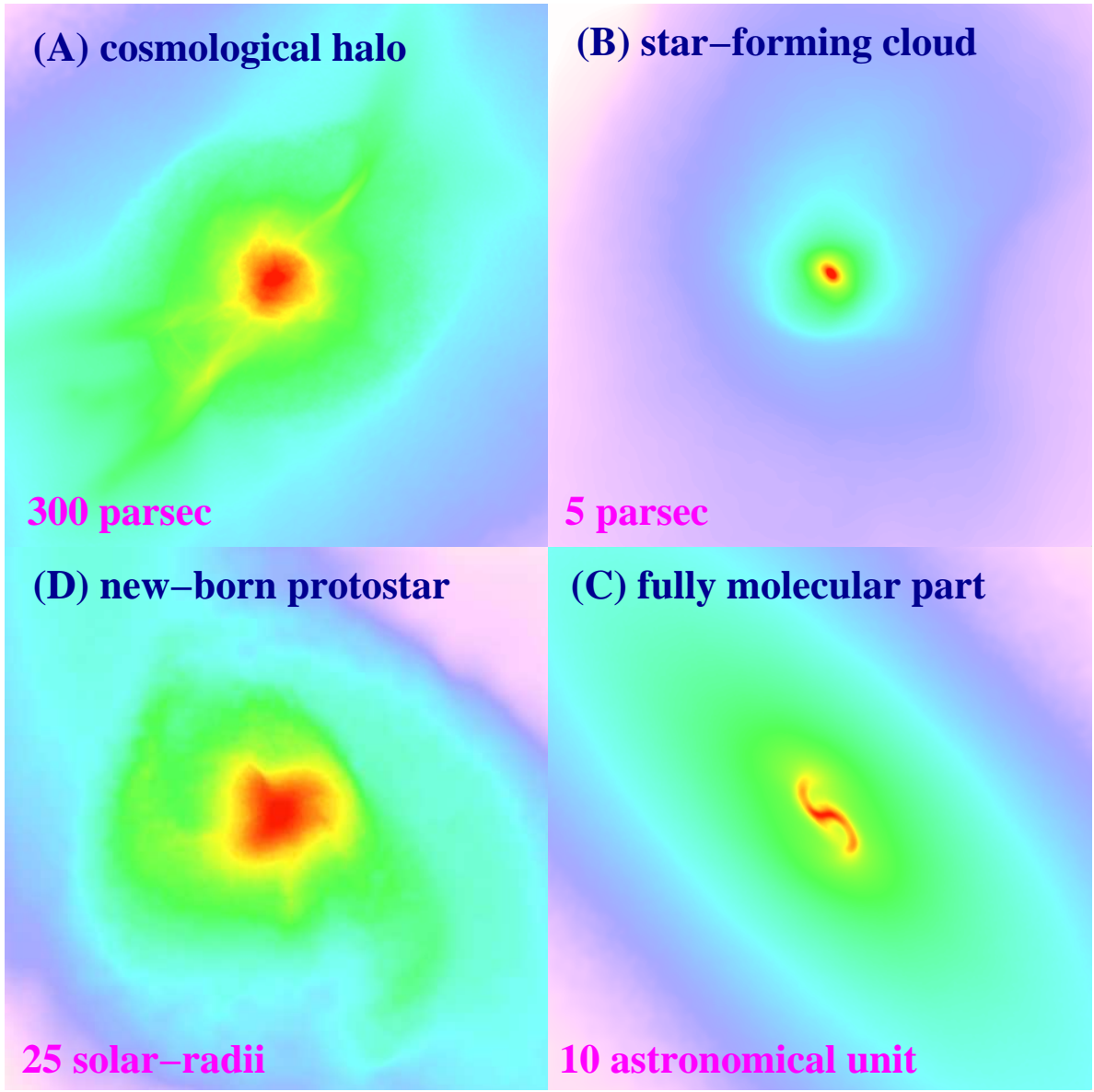}
\caption{Projected gas distribution around a primordial protostar\cite{YOH08}.
Shown are, 
(A) the large-scale gas distribution around the cosmological minihalo
(300 pc on a side), (B) a self-gravitating, star-forming cloud (5 pc on a side), 
(C) the central part of the fully molecular core (10 astronomical units on a side),
and (D) the final protostar (25 solar-radii on a side).
\label{fig:protostar}}
\end{figure*}

Although the mass triggering the first runaway collapse is
well-determined, it provides only a rough guess of the mass of the
star(s) to be formed. Standard star formation theory predicts that a
tiny protostar forms first and subsequently grows by accreting the
surrounding gas to become a massive star.  Indeed, the highest
resolution simulations of first-star formation verify that this also
occurs cosmologically\cite{YOH08} 
(see Fig. \ref{fig:protostar}).  
However, the ultimate mass of the
star is determined both by the mass of the cloud out of which the star
forms and by a number of feedback processes that occur during the
evolution of the protostar. In numerical simulations, the final mass
of a Pop~III star is usually estimated from the density distribution
and velocity field of the surrounding gas when the first protostellar
fragment forms, but this may well be inaccurate even in the absence of
protostellar feedback.  While protostellar feedback effects are well
studied in the context of the formation of contemporary
stars\cite{MO07}, they differ in several important respects in
primordial stars\cite{McK08}.

First, primordial gas does not contain dust grains.  As a result,
radiative forces on the gas are much weaker. 
Second, it is
generally assumed that magnetic fields are not important in primordial
gas because, unless exotic mechanisms are invoked, the amplitudes of
magnetic fields generated in the early Universe are so small that they
never become dynamically significant in primordial star-forming
gas\cite{Ryu08}.  Magnetic fields have at least two important effects
in contemporary star formation: they reduce the angular momentum of
the gas out of which stars form and they drive powerful outflows
that disperse a significant fraction of the parent cloud. It is
likely that the pre-stellar gas has more angular momentum in the
primordial case, and this is borne out by cosmological
simulations. Third, primordial stars are much hotter than contemporary
stars of the same mass, resulting in significantly greater ionising
luminosities\cite{Schaerer02}.

\begin{figure*}[hptb]
\plotone{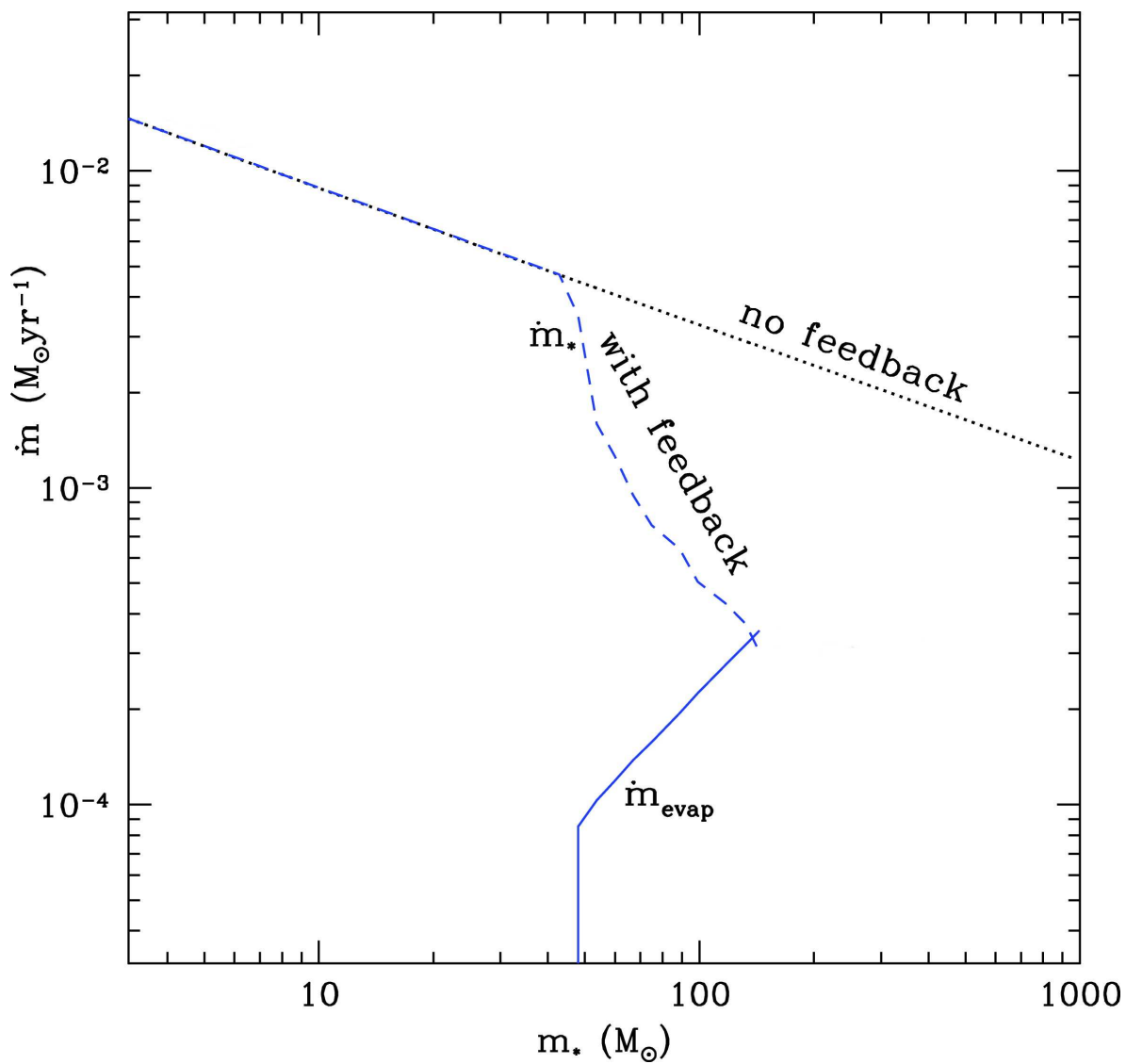}
\caption{Feedback limited accretion\cite{McK08}. The accretion
rate vs. protostellar mass is shown in the cases of ``no feedback'' and ``with
feedback''. Even as an H~{\sc II} region
is built up, accretion continues through an accretion disk, which is
eventually destroyed via photoevaporation. Also shown is the corresponding rate.
The intersection of the two curves determines the final Pop~III mass.
\label{fig:evap}}
\end{figure*}

State-of-the-art numerical simulations of the formation of the first
(Pop~III.1) stars represent a computational {\it tour de force}, in
which the collapse is followed from cosmological (comoving Mpc) down
to protostellar (sub astronomical-unit) scales, revealing 
the entire formation process of a protostar. 
However, further growth of the protostar cannot be followed accurately
without implementing additional radiative physics. 
For now, inferring the
subsequent evolution of the protostar requires approximate analytic
calculations. By generalising a theory for contemporary massive star
formation\cite{mkt02}, it is possible to approximately reproduce the
initial conditions found in the simulations and to then predict the
growth of the accretion disk around the star\cite{Tan04}.  Several
feedback effects determine the final mass of a first star\cite{McK08}:
Photodissociation of H$_2$ in the accreting gas reduces the cooling
rate, but does not stop accretion. Lyman-$\alpha$ radiation pressure
can reverse the infall in the polar regions when the protostar grows
to $20-30 M_{\odot}$, but cannot significantly reduce the accretion
rate. The expansion of the H~{\sc II} region produced by the large
flux of ionising radiation can significantly reduce the accretion rate
when the protostar reaches $50-100 M_{\odot}$, but accretion can
continue in the equatorial plane. Finally, photoevaporation-driven
mass loss from the disk\cite{Hol94} stops the accretion and fixes the
mass of the star (see Fig.~\ref{fig:evap}). The final mass depends on the entropy and
angular momentum of the pre-stellar gas; for reasonable conditions,
the mass spans $60-300 M_{\odot}$.

A variety of physical processes can affect and possibly substantially alter
the picture outlined above.
Magnetic fields generated through the magneto-rotational instability 
may become important in the proto-stellar disk\cite{TanB03}, 
although their strength is uncertain,
and may play an important role in the accretion phase\cite{Machida07}.
Cosmic rays and other external ionisation sources, if they existed in the 
early Universe, could significantly affect the evolution of primordial
gas\cite{Stacy07}.
A partially ionised gas cools more efficiently because the abundant electrons
promote ${\rm H}_2$ formation. Such a gas cools to slightly lower temperatures
than a neutral gas can, 
accentuating the fractionation of D into HD so that
cooling by HD molecules becomes 
important\cite{Yoshida07}$^-$\cite{McGB08}.

\begin{figure*}[hptb]
\epsscale{0.4}\plotone{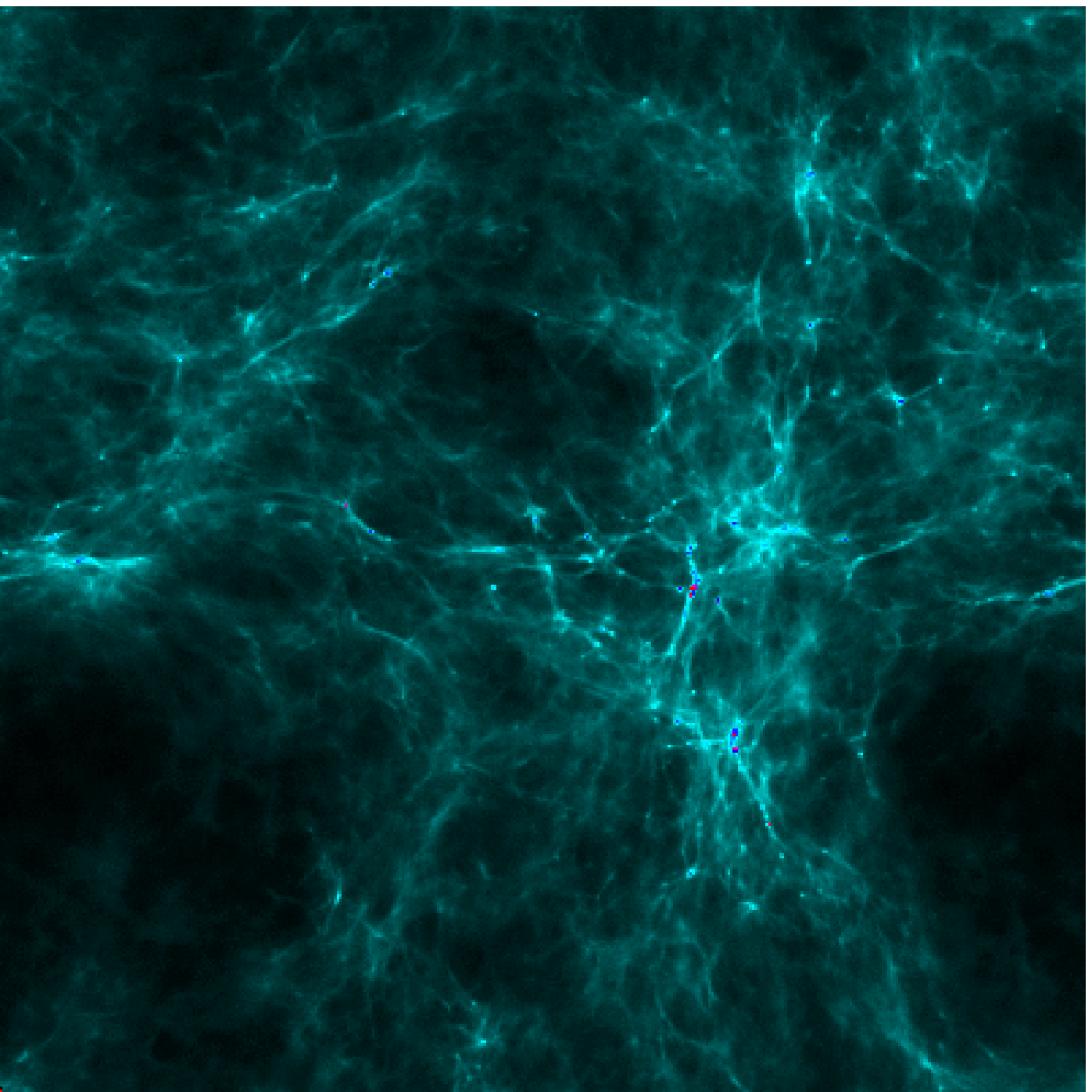}
\plotone{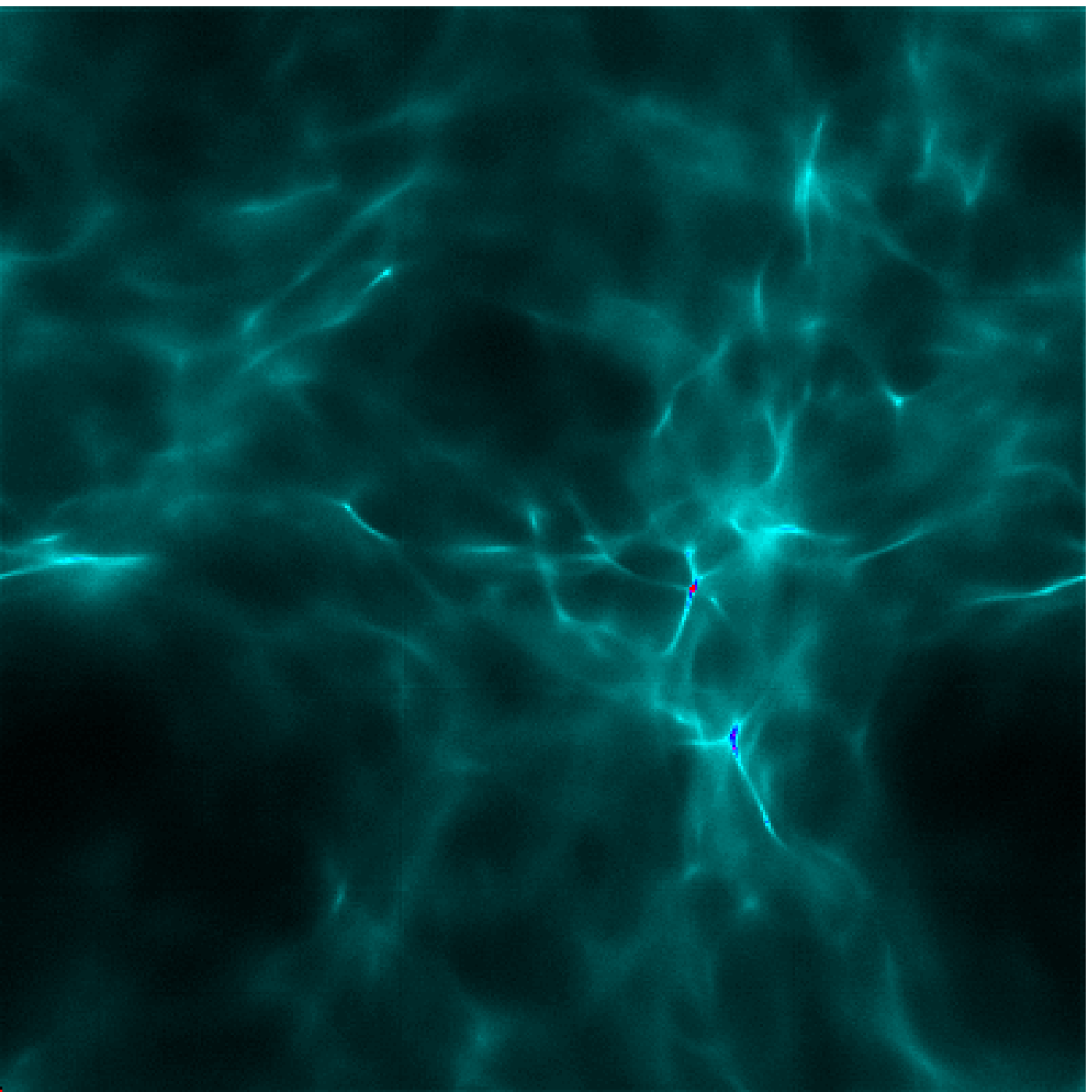}
\caption{Dark matter properties and early star formation\cite{YSHS03}.
Projected gas distribution in CDM ({\it left}) and warm dark matter (WDM)
simulations ({\it right}) at $z=20$. 
If the power in the primordial density spectrum is reduced on small
scales, the first stars will form much later than in the standard
CDM-based scenario. If the dark matter is warm, having a substantial
velocity dispersion, density perturbations on small length scales
are smoothed. The hierarchy of structure formation is then truncated
at a corresponding mass scale, and the first cosmological
objects could be more massive than $10^6 M_{\odot}$. For the case of
light warm dark matter\cite{GaoT07}, gas collapses into filaments,
which might then fragment into multiple stellar cores.  The abundance
of star-forming halos is significantly reduced in this model.
\label{fig:darkmatter}}
\end{figure*}

More significant modifications to the standard model result if the
properties of the dark matter are different from
those assumed above
(see Fig.~\ref{fig:darkmatter}).
A key assumption in the standard model is that the dark matter interacts
with the baryons only via gravity. However, dark matter can indirectly
affect the dynamics of a pre-stellar gas.
A popular candidate for CDM is the neutralino, for which the
self-annihilation cross-section is large. Neutralino dark matter is thus
expected to pair-annihilate in very dense regions, producing high-energy 
particles such as pions and electron-positron pairs and high-energy photons.
These annihilation products may effectively heat collapsing 
primordial gas clouds when the density is sufficiently high,
thereby arresting the collapse\cite{Spolyar08}.
Calculation of the structure of stars with dark-matter annihilation suggest that they
can undergo a phase of evolution in which they have temperatures of $4000-10^4$~K, well 
below those for conventional Pop~III stars\cite{Freese08}$^,$\cite{Iocco08}.
The magnitude of this effect depends sensitively on details
such as the dark matter concentration and the final products of neutralino
annihilation.
Furthermore, calculations to date have assumed spherical symmetry,
whereas it is possible that the angular momentum of both the baryons
(which leads to the formation of an accretion disk\cite{Tan04}) and of the dark
matter could significantly impede the
buildup of the high dark-matter densities required to power the stellar
luminosity via dark-matter  annihilation.
Nevertheless, if neutralinos are detected in the appropriate mass
range\cite{LHC}, early star formation models 
may need to include the effect of dark matter annihilation.\\

\noindent{\Large \bf Feedback from the first stars}

Some of the feedback processes described above that affect the
formation of individual stars also impact primordial star formation on
large scales.  The enormous fluxes of ionising radiation and
H$_2$-dissociating Lyman-Werner (LW) radiation emitted by massive
Pop~III stars\cite{Schaerer02}$^,$\cite{BrommK01} dramatically
influence their surroundings, heating and ionising the gas within a
few kpc of the progenitor and destroying the H$_2$ within a somewhat
larger region\cite{Kitayama04}$^,$\cite{Yoshida07}$^,$\cite{Whalen04}$^-$\cite{Abel07}.
Moreover, the LW radiation emitted by the first stars could propagate
across cosmological distances, allowing the build-up of a pervasive LW
background radiation field\cite{Ciardi00}$^,$\cite{Haiman00}.
The impact of radiation from the first stars on their local surroundings 
has important implications for the numbers and types of Pop~III stars that 
form. The photoheating of gas in the minihalos hosting Pop~III.1 stars 
drives strong outflows, lowering the density of the gas in the minihalos and 
delaying subsequent star formation by up to $100~\rm{Myr}$\cite{JGB07}.
Furthermore, neighbouring minihalos may be photoevaporated, delaying star 
formation in such systems as well\cite{Susa06}$^-$\cite{Whalen07}. The 
photodissociation of molecules by LW photons emitted from local star-forming 
regions will, in general, act to delay star formation by destroying the 
main coolants that allow the gas to collapse and form stars\cite{Machacek01}.

\begin{figure*}[hptb]
\epsscale{0.9}\plotone{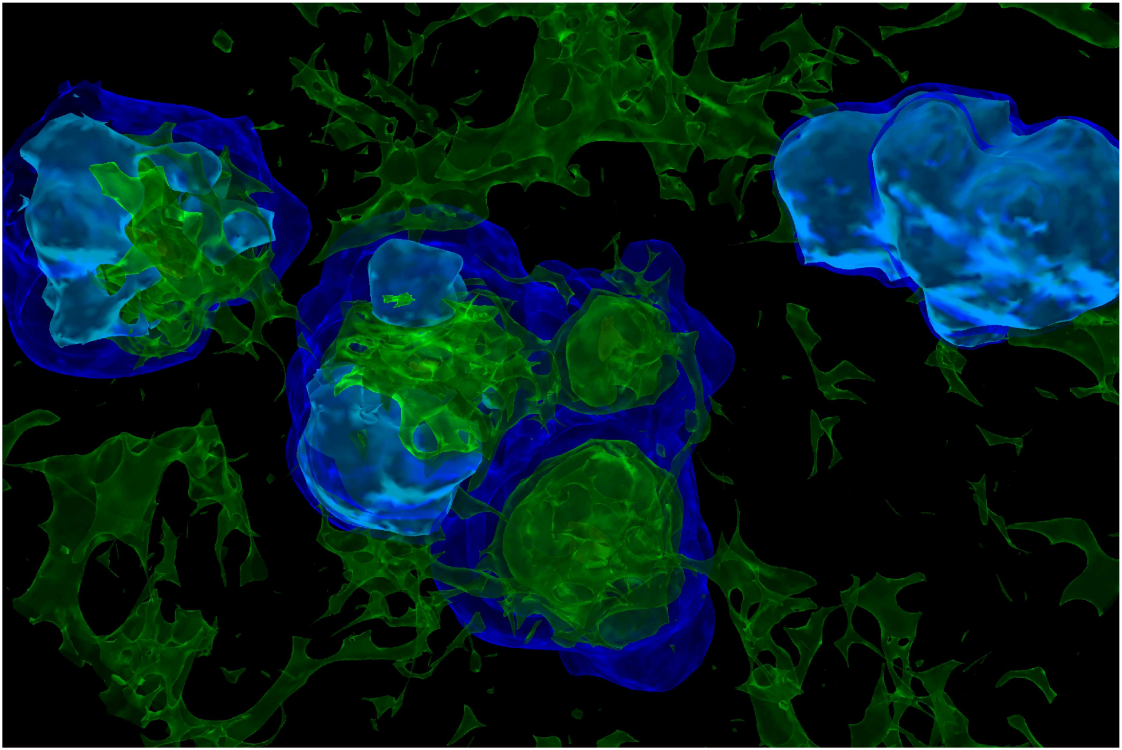}
\caption{Radiative feedback around the
first stars\cite{JGB07}. Ionized bubbles are shown in blue, and
regions of high molecule abundance in green. The large residual free electron
fraction inside the relic H~{\sc II} regions, left behind after the central star has died, rapidly catalyzes the
re-formation of molecules. The abundance of HD molecules 
allows the primordial gas to cool to the 
temperature of the CMB, possibly leading to the formation of Pop~III.2 stars 
after these regions have re-collapsed so that gas densities are sufficiently
high again for gravitational instability to occur\cite{Yoshida07b}. The latter
process takes of order the local Hubble time, thus imposing a
$\sim 100$~Myr delay in star formation.
The relatively high molecule abundance in relic H~{\sc ii} regions, 
along with their increasing 
volume-filling fraction, leads to a large optical depth to LW photons over 
physical distances of the order of several kpc\cite{JGB07}.
The development of a high 
optical depth to LW photons over such short length-scales, combined
with a rapidly increasing volume filling fraction of relic H~{\sc ii}
regions, suggests that the 
optical depth to LW photons over cosmological scales may be very high, 
acting to suppress the build-up of a background LW radiation field, 
and mitigating negative feedback on star formation\cite{Johnson08}.
Note the strongly clustered nature of early star formation. 
\label{fig:bubbles}}
\end{figure*}

The photoionisation of primordial gas, however, 
can also stimulate star formation by fostering the 
production of abundant molecules within the relic H~{\sc ii} 
regions surrounding the remnants of Pop~III.1 stars\cite{Abel07}$^,$\cite{JGB07}$^,$\cite{Ricotti01}$^,$\cite{OhH02}
(see Fig.~\ref{fig:bubbles}).
It is still debated whether this radiative feedback is positive or
negative in terms of its 
overall impact on the 
cosmic star
formation rate\cite{Ferrara98}. However, some robust conclusions
have emerged from the recent simulations. First,
the LW feedback is much less `suicidal' than was originally thought\cite{HRL97}.
It is now believed that star formation in neighbouring minihalos is not
completely suppressed, but merely delayed. Second, the ionising radiation
from the first stars is initially very disruptive 
because it substantially
decreases the density in the host minihalo. This effect leads to
the substantial gap between the formation of the first and second
generations of stars. In each region of space, the drama of `first light' thus occurred in two
clearly separated stages.

 Most of the work on the evolution of Pop~III stars and on the SNe they
produce has been based on the assumption that the stars are not rotating\cite{HegerW02}.
For initial stellar masses in the range 
$25 M_\odot\la M_*\la 140 M_\odot$ and $M_*\ga 260 M_\odot$, Pop~III stars end their lives by
collapsing into black holes with relatively little ejection of heavy elements. Pop~III stars
in the range $140-260 M_\odot$ explode as pair-instability supernovae 
(PISNe), which disrupt the entire progenitor, with explosion energies ranging from 
$10^{51}-10^{53}~\rm{erg}$, and nucleosynthetic yields, defined as the
heavy element mass fraction, up to $0.5$.
Such SNe exhibit an odd-even effect in the nuclei produced that is much greater than
observed in any star to date, and as a result they cannot make a significant contribution
to the metals observed in very low-metallicity stars today\cite{Tum06}.
On the other hand, the PISN signature may exist in a tiny fraction of the stars
with intermediate metallicity ($\sim 0.01 Z_{\odot}$), because the enrichment
from even a single PISN already endows the surrounding material with
heavy elements to levels that
are above the regime typically probed by surveys of metal-poor stars\cite{Karlsson08}.

The first stars may have been born rapidly rotating, however, and rotation can
entirely modify these results\cite{Mae07}.
For sufficiently high rotation rates, rotationally induced
mixing is able to render the cores chemically homogeneous; mixing of heavy elements
to the surface in the late stages of evolution can lead to substantial mass loss. If
the cores maintain a sufficiently high rotation at the time of the SN, it is
possible to produce a long gamma-ray burst (GRB) or a jet-induced energetic supernova/hypernova\cite{YoonL05}$^,$\cite{WoosleyH06},
with significant effects on
the abundances of the ejected metals\cite{nom08}. Large uncertainties
remain in the evolutionary calculations owing to the effects of dynamo-generated
magnetic fields.

 The strong mechanical and chemical feedback effects exerted by 
explosions of Pop~III stars have been investigated with a number of detailed
calculations\cite{GJBK07}$^-$\cite{Whalen08}.
The key question is how the initially metal-free Universe was enriched
with the first heavy chemical elements\cite{Tornatore07}.
Recently, it has become feasible to address this process with realistic
three-dimensional simulations that start from cosmological initial
conditions, and that resolve the detailed physics of the SN blast wave
expansion\cite{GJBK07}$^,$\cite{WA08}. These simulations
have shown that early enrichment is very inhomogeneous, as the low-density
voids are enriched before any metals can reach into the denser filaments
and virialised halos\cite{CenR08}.\\ 

\noindent{\Large \bf Assembly of the first galaxies}

The characteristic mass of the first star formation sites has been determined 
to be $\sim 10^6 M_{\odot}$\cite{Teg97}$^,$\cite{YAHS03}, whereas the critical
mass for hosting the formation of the first galaxies is still not known with
any certainty. A promising theoretical {\it Ansatz} is to explore 
atomic cooling halos, with $\sim 10^8 M_{\odot}$ and
virial temperatures greater than $\sim 10^{4}$~K so that atomic line cooling is efficient,
as their formation sites\cite{Wise07}$^,$\cite{Greif08}. 
The simulations, starting from cosmological initial conditions, are just now
approaching the resolution and physical realism to investigate whether atomic cooling
halos fulfill the criteria for a first galaxy as defined above.
Quite generically, in such models, the first generation of stars 
forms before galaxies do, and 
{\it feedback effects from the first stars are expected to play a key role in 
determining the initial conditions for the formation of the first galaxies.}
While substantial uncertainties in the overall formation efficiency 
of the first stars still remain, it is possible, 
and perhaps probable, that most first galaxies
hosted at least one primordial star earlier\cite{Johnson08}.
If the early generation stars were massive, $\ga 10 M_{\odot}$,
the feedback effects described in the previous section would shape
the conditions for subsequent star-formation in the region.

The gas expelled by the H~{\sc II} regions and SNe of the first stars
is too hot and diffuse to allow further star formation until it had time
to cool, as well as to reach high densities again in the course of being
reincorporated in a growing dark matter halo. Both cooling and re-collapse
occur rather slowly, thus rendering star-formation intermittent in the early
formation phase of the first galaxies.
Analytic models\cite{YBH04} and detailed numerical 
simulations\cite{JGB07}$^,$\cite{Yoshida07b} both show
that the gas re-incorporation time is as long as a hundred million years,
roughly corresponding to the dynamical time for a first-galaxy halo to be assembled.

Chemical enrichment by the first SNe
is among the most important processes in the formation of the first galaxies.
Efficient cooling by metal lines and dust thermal emission regulate the
temperature of Pop~II star-forming regions in the first galaxies.
The concept of a `critical metallicity' has been introduced to characterise the 
transition of the star-formation mode from predominantly high-mass, Pop~III or Pop~II,
to low-mass Pop~II stars\cite{BrommF01}. 
However, this critical gas metallicity is still
poorly determined.
It is not even clear if there exists such a sharp transition. 
Some studies show that even a slight quantity of metals in a gas may be enough
to change the gas thermal evolution significantly\cite{Omukai05}, 
whereas others argue that the cooling efficiency at low densities\cite{bl04}
is crucial and is significantly enhanced only
above one ten-thousandth of solar-metallicity ($Z \ga 10^{-4} Z_{\odot}$).
Since the enrichment from even a single PISN by a very massive Pop~III
star likely leads to metallicities of $Z > 10^{-2} Z_{\odot}$\cite{GJBK07}, 
well in excess of {\it any} predicted value for the critical metallicity, 
these arguments
might be somewhat academic. The characteristic mass of prestellar gas clumps is likely
determined by a number of physical processes (e.g., turbulence and, possibly, dynamo-amplified
primordial magnetic fields)
other than radiative cooling. The overall effect of gas metallicity on star-formation may well be 
limited\cite{Jappsen08}.

\begin{figure*}[hptb]
\plotone{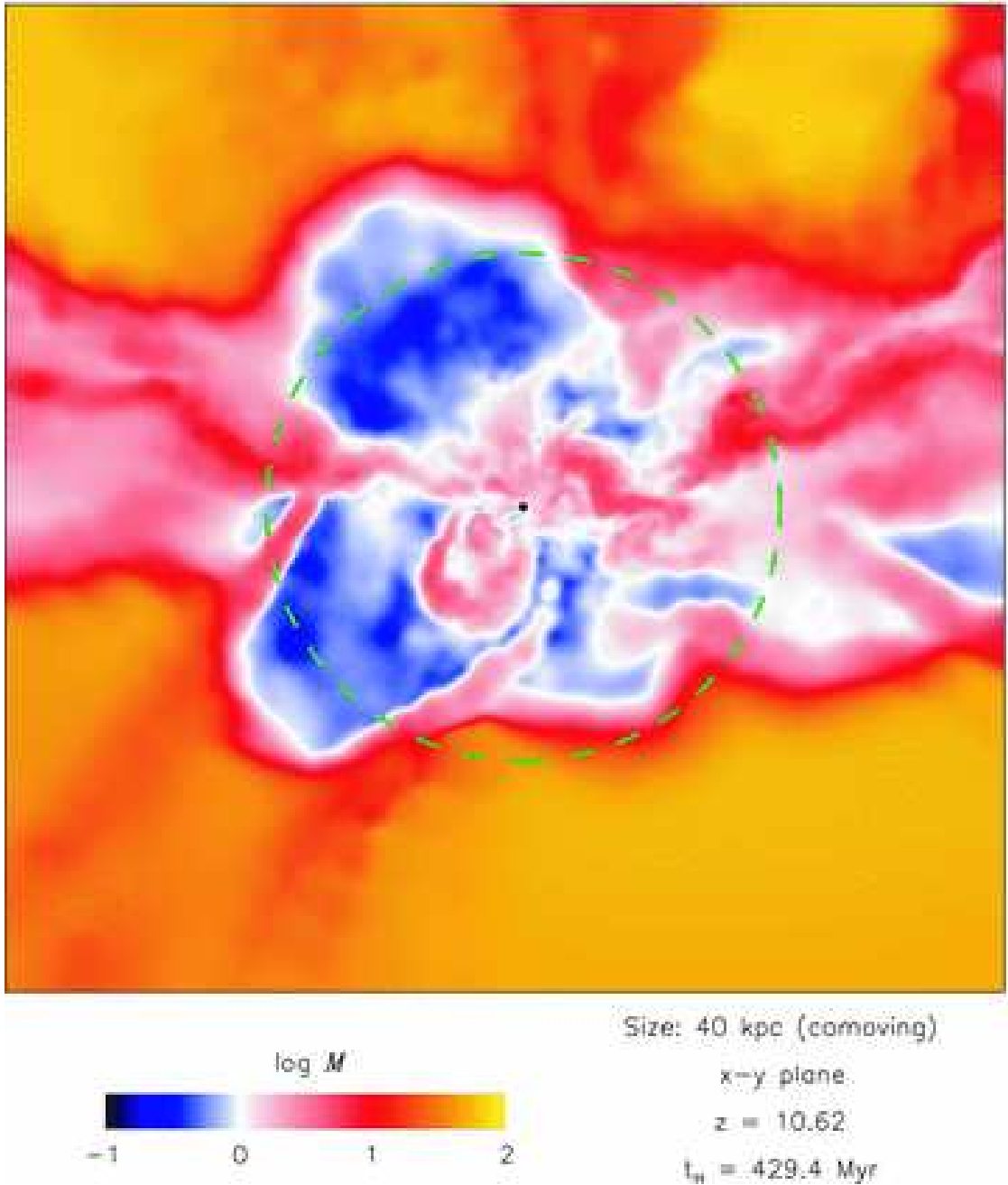}
\caption{Turbulence inside the first galaxies\cite{Greif08}. Shown is the Mach
number in a slice through the central 40~kpc (comoving) of the galaxy.
The dashed line denotes the virial radius of $\simeq 1$~kpc. The Mach
number approaches unity at the virial shock, where the accreted gas
is heated to the virial temperature. Inflows of cold gas along filaments
are supersonic by a factor of $\simeq 10$, resulting in strong turbulent
flows in the galactic center.
\label{fig:turbulence}}
\end{figure*}

Recent cosmological simulations have demonstrated that star formation
inside the first galaxies is strongly influenced
by gravitationally-driven supersonic turbulence
that was generated during the virialisation 
process\cite{WA08}$^,$\cite{Wise07}$^,$\cite{Greif08}.
This is in marked difference to
the rather quiescent, quasi-hydrostatic situation in minihalos (see Fig.~\ref{fig:turbulence}).
It thus appears possible that the first galaxies harbour the
first stellar clusters,
if present-day star formation offers any guide here, where
it is widely believed that gravo-turbulent fragmentation is responsible 
for shaping the initial mass function (IMF)\cite{MO07}$^,$\cite{MLK04}.
It is an open question as to whether the first galaxies could harbour the first globular clusters,
which are the oldest star clusters known.\\

\noindent{\Large \bf Future empirical probes}

Studying the formation of the first stars and galaxies will be at the frontier of 
astronomy and cosmology in the next decade. Astronomers will muster
a comprehensive arsenal of observational probes.
The most prominent among these 
concern the CMB optical depth to Thomson
scattering\cite{Fan06}$^-$\cite{Holder03}, 
the near-IR background\cite{Kashlinsky05},
high-redshift GRBs\cite{LR00}$^-$\cite{Totani06},
the possibility of scrutinising 
the nature of the first stars by metals found in the oldest Galactic 
halo stars, dubbed `stellar archaeology'\cite{fjb07}$^,$\cite{Iwamoto05},
and various facilities now
being deployed to map reionisation using the redshifted 21 cm line
of neutral hydrogen\cite{Furlanetto06}$^-$\cite{mcquinn07}. 
The {\it James Webb Space Telescope (JWST)} will carry out a number of observations designed to test
key assumptions of our current theory of the first stars and
galaxies\cite{Gardner06}.
How could the existence of massive Pop~III stars be unambiguously inferred?
The most clear-cut diagnostic is the ratio of recombination lines
emitted from the H~{\sc II} regions around single Pop~III stars,
or clusters thereof, to be measured with ultra-deep near-IR and mid-IR spectroscopy.
Due to the high effective temperature of the
Pop~III stellar continuum, $\sim 10^5$~K, strong He~{\sc II} $\lambda$~1640
line emission is predicted, with a ratio compared to Lyman-$\alpha$ that
is one to two orders of magnitude larger than for normal stars\cite{BrommK01}.
A second crucial observational campaign aims at a census of very high-$z$
SNe\cite{Mackey03} through deep broadband near-IR imaging.
One key objective is to search for possible
PISN events, which would clearly stand out owing to their extreme intrinsic
brightness, as well as their very long durations, a few years in the
observer frame\cite{Scannapieco05}.
The goal of making 
useful predictions for the high-redshift frontier is now clearly drawing within 
reach, and the pace of progress is likely to be rapid.\\
\newline

\vskip 0.2in

\noindent

{\bf Acknowledgments} The authors are grateful for the hospitality of KITP,
University of California Santa Barbara.
This work was supported in part by NSF and NASA.
\vskip 0.2in

\noindent

{\bf Correspondence} and requests for materials should be addressed to
V.B.\\
(email: vbromm@astro.as.utexas.edu) or 
N.Y. (email: naoki.yoshida@ipmu.jp).

\clearpage

\clearpage

\clearpage

\clearpage

\clearpage

\clearpage

{

\noindent{\Large \bf Box: Definitions and Terminology}

\noindent
We here establish a convention for terminology to be used in this review.
Pop~III stars are those that
initially contain no elements heavier than helium ('metals' in the
parlance of astronomers) other than the lithium produced in the Big
Bang. Such stars can be divided into first generation stars
(Pop~III.1), which form from initial conditions determined entirely by
cosmological parameters, and second generation stars (Pop~III.2),
which originate from material that was influenced by earlier star
formation\cite{McK08}. According to theory,
Pop~III.1 stars formed when almost completely neutral primordial gas
collapsed into dark-matter minihalos, whereas one important class of
Pop~III.2 stars formed from gas that was photoionised prior to the
onset of gravitational runaway collapse\cite{Yoshida07}.  Simply put,
Pop~III.1 stars are locally the very first luminous objects, whereas
Pop~III.2 stars are those metal-free stars formed from gas that was
already affected by previous generations of stars. Pop~II
stars have enough metals to affect their formation and/or their
evolution.
Such stars are classified\cite{beers05} according to their
iron/hydrogen ratio as extremely metal poor (EMP) for metallicities
$10^{-4}<Z/Z_{\odot} <10^{-3}$, ultra-metal poor (UMP) for
$10^{-5}<Z/Z_{\odot} <10^{-4}$, and hyper-metal poor (HMP)
$10^{-6}<Z/Z_{\odot} <10^{-5}$.
Because we know so little about the first galaxies, it is difficult to
establish a precise terminology for them. A galaxy is a system of many
stars and gas that is gravitationally bound in a dark matter halo. We
define a {\it first galaxy} as one comprised of the very first system
of stars to be gravitationally bound in a dark matter halo. Such stars
could be Pop~III or Pop~II stars with very low metallicities---EMP or
below according to recent numerical
simulations\cite{GJBK07}$^,$\cite{WA08}. The gas in such galaxies
should have similarly low metallicities. Current
theory predicts that first generation Pop~III stars (Pop III.1) are
formed in isolation in minihalos and therefore will {\it not} be in
galaxies.

\end{document}